\documentclass[twocolumn,superscriptaddress,prb]{revtex4}
\usepackage{amsmath}
\usepackage{amssymb}
\usepackage{graphicx}
\usepackage{bm}
\usepackage{color}			
\usepackage[normalem]{ulem} 

\def\Vec#1{{\bf #1}}
\def\GVec#1{\mbox{\boldmath $#1$}}

\begin{document}


\title{Hofstadter butterfly and the quantum Hall effect in twisted double bilayer graphenes}

\author{J. A. Crosse}
\affiliation{Arts and Sciences, New York University Shanghai, 1555 Century Ave, Pudong, Shanghai, 200122, China.}
\affiliation{NYU-ECNU Institute of Physics at NYU Shanghai, 3663 Zhongshan Road North, Shanghai, 200062, China.}
\author{Naoto Nakatsuji}
\affiliation{Department of Physics, Osaka University, Osaka 560-0043, Japan}
\author{Mikito Koshino}
\affiliation{Department of Physics, Osaka University, Osaka 560-0043, Japan}
\author{Pilkyung Moon}
\email{pilkyung.moon@nyu.edu}
\affiliation{Arts and Sciences, New York University Shanghai, 1555 Century Ave, Pudong, Shanghai, 200122, China.}
\affiliation{NYU-ECNU Institute of Physics at NYU Shanghai, 3663 Zhongshan Road North, Shanghai, 200062, China.}
\affiliation{Department of Physics, New York University, 726 Broadway, New York, NY 10003, USA}
\affiliation{State Key Laboratory of Precision Spectroscopy, East China Normal University, Shanghai 200062, China}

\date{\today}

\begin{abstract}

We study the energy spectrum and quantum Hall effects of the twisted double bilayer graphene in uniform magnetic field.
We investigate two different arrangements, AB-AB and AB-BA, which differ in the relative orientation
but have very similar band structures in the absence of a magnetic field.
For each system, we calculate the energy spectrum and quantized Hall conductivities at each spectral gap
by using a continuum Hamiltonian that satisfies the magneto-translation condition.
We show that the Hofstadter butterfly spectra of AB-AB and AB-BA stackings differ significantly,
even though their zero magnetic field band structures closely resemble;
the spectrum of AB-AB has valley degeneracy, which can be lifted by applying interlayer potential asymmetry,
while the spectrum of AB-BA has no such degeneracy in any case.
We explain the origin of the difference
from the perspectives of lattice symmetry and band topology.

\end{abstract}

\maketitle

\section{Introduction}

\begin{figure}
	\centering
	\includegraphics[width=0.8\columnwidth]{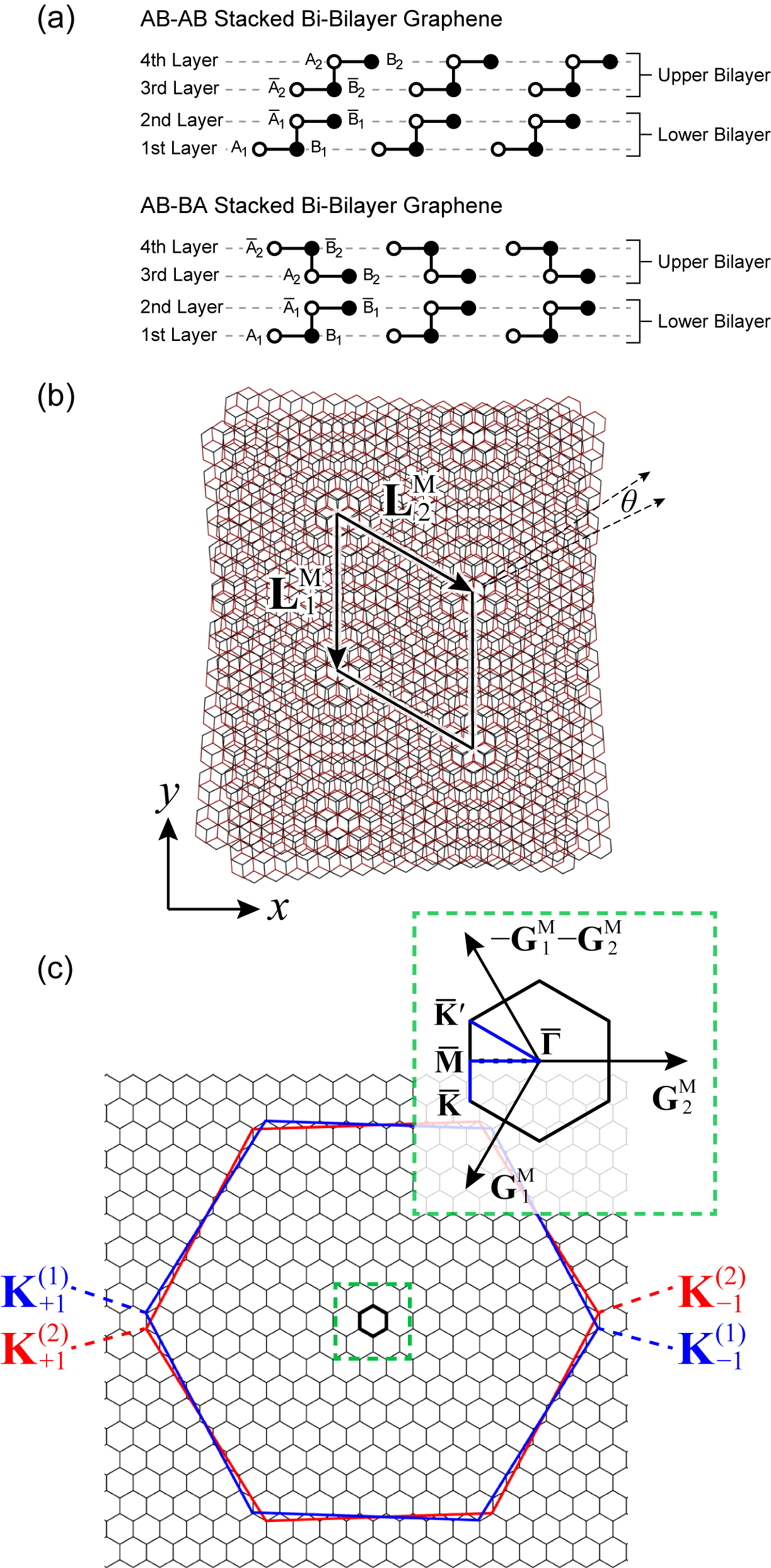}
	\caption{(a) Cross-section through TDBG showing the atomic alignment of the AB-AB and AB-BA stacking, respectively. (b) The atomic structure of TDBG. (c) Brillouin zone folding in TDBG. The large red and blue hexagons represent the Brillouin zone of the two graphene bilayers. The small hexagons represent the moir\'{e} Brillouin zone of TDBG. Inset shows the high-symmetry points of the moir\'{e} Brillouin zone.}
	\label{fig1}
\end{figure}

With the recent observation of unconventional superconductivity in twisted bilayer graphene \cite{Cao2018, Cao2018b, Yankowitz2019}, two-dimensional van der Waals heterostructures have become the preferred system for studying the phases of strongly correlated electrons.
In general, the band structures and electron interaction strength in twisted bilayer graphene are determined by the twist angle between the layers,
which is fixed at the moment of fabrication.
In contrast, twisted double bilayer graphene (TDBG), a pair of AB-stacked graphene bilayers stacked at an angle $\theta$ with respect to each other [Figs.\,\ref{fig1}(a) and \ref{fig1}(b)], has been shown to display electrically tunable flat bands at various twist angles. Hence, TDBG could potentially serve as a better platform for device design, as valleys, layers, and gap size can be controlled via an applied bias \cite{Rickhaus2019gap,de2020combined}, and so far displayed novel features such as superconductivity, correlated insulating states and ferromagnetic order \cite{Burg2019correlated,liu2019spin,shen2019observation,cao2019electric,adak2020tunable,he2020tunable}. 
To date, there have only been a few theoretical studies of the electronic properties of TDBG. These mostly investigated the electronic band structures and the response to an interlayer asymmetric potential \cite{Koshino2019,Liu2019quantum,Chebrolu2019,Choi2019,haddadi2019moir,culchac2020flat}, although a there have also been some studies on
ferromagnetic and superconducting order \cite{Lee2019theory,wu2019ferromagnetism,samajdar2020microscopic} and their instability \cite{scheurer2019pairing}.

TDBG has two different configurations,
AB-AB stacking and AB-BA stacking,
which are obtained by rotating one of the bilayers by $180^\circ$ [Fig.\,\ref{fig1}(a)].
The band structures of these two configurations are almost identical.
In terms of the band topology, however,
their topological character, such as the valley Chern number in each gap, differ significantly \cite{Koshino2019,Liu2019quantum}.
This motivates us to explore the possible signatures of this topological difference.

In this work, we calculate the Landau level spectrum and Hall conductivities of TDBG in both the AB-AB and AB-BA configurations.
We show that the two configurations of TDBG, owing to their different topologies,
exhibit totally different energy spectra in magnetic fields $B$
in spite of the similarity between their $B=0\,\mathrm{T}$ band structures.
The valley degeneracy of the Landau levels is closely related to the symmetry of the systems and the valley Chern numbers of the $B=0\,\mathrm{T}$ bands.
We found that the Landau levels of the intrinsic AB-AB stacked TDBG
are valley degenerate, while the degeneracy can be lifted,
for example, by applying interlayer potential asymmetry.
In contrast, the Landau levels of the AB-BA stacked TDBG have no valley degeneracy in any case.
A non-zero valley Chern number generally correlates with a significant difference in the spectrum between the $K$ and $K'$ monolayer valleys,
because 
the different Chern numbers in $K$ and $K'$ lead to the different dependences 
of the electron density below a gap 
as a function of magnetic field.



The paper is organized as follows.
Sections \ref{sec:atomic_structure}, \ref{sec:effective_continuum_model}, and \ref{sec:electronic_structure_in_a_magnetic_field}
present the atomic structures, effective continuum Hamiltonian,
and the magnetic Bloch functions of TDBG.
We discuss the symmetries of the lattice and Hamiltonian of both the AB-AB and AB-BA configurations in Sec.\,\ref{sec:symmetry_argument},
and investigate the Landau level spectrum and Hall conductivities in Sec.\,\ref{sec:energy_spectrum},
where we discuss in detail about the similarity and difference between the two configurations.
We conclude in Sec.\,\ref{sec:conclusion}.

\section{Theoretical Methods}

\subsection{Atomic Structure}
\label{sec:atomic_structure}

We choose our coordinate system such that the lattice vectors of the unrotated graphene bilayers are given by $\mathbf{a}_{1} = a(1,0)$ and $\mathbf{a}_{2} = a(1/2,\sqrt{3}/2)$ with the graphene lattice constant $a=0.246\,\mathrm{nm}$. After rotation, the lattice vectors in each layer read $\mathbf{a}_{i}^{(l)} = R(\mp\theta/2) \mathbf{a}_{i}$ where $R(\theta)$ is the rotation matrix and $l \in 1, 2$ refers to rotations of $\mp\theta$, respectively.
Accordingly, the unrotated reciprocal lattice vectors are given by $\mathbf{a}_{1}^{\ast} = (2\pi/a)(1,-1/\sqrt{3})$ and $\mathbf{a}_{2}^{\ast} = (2\pi/a)(0,2/\sqrt{3})$ and the rotated reciprocal lattice vectors in each layer by $\mathbf{a}_{i}^{\ast(l)} = R(\mp\theta/2) \mathbf{a}_{i}^{\ast}$. The Dirac points of the two graphene layers are located at $\mathbf{K}_{\xi}^{(l)}=-\xi[2\mathbf{a}_{1}^{\ast(l)}+\mathbf{a}_{2}^{\ast(l)}]/3$, with $\xi=\pm 1$ labeling the $K$ and $K'$ valleys. When a small twist angle is introduced the lattices of the two layers form a moir\'{e} interference pattern leading to an dramatic increase in the size of the material's unit cell and a correspondingly large reduction in the size of the Brillouin zone [see Fig.\,\ref{fig1}(c)]. The reciprocal lattice vectors for the moir\'{e} Brillouin zone are given by $\mathbf{G}_{i}^{\mathrm{M}}=\mathbf{a}_{i}^{\ast(1)}-\mathbf{a}_{i}^{\ast(2)}$, from which the real space moir\'{e} lattice vectors can be found via $\mathbf{G}_{i}^{\mathrm{M}}\cdot\mathbf{L}_{j}^{\mathrm{M}} = 2\pi$. The moir\'{e} lattice constant is given by $L_{\mathrm{M}} = a/[2\sin(\theta/2)]$ and the area of the moir\'{e} unit cell by $A=|\mathbf{L}_{1}^{\mathrm{M}}\times\mathbf{L}_{2}^{\mathrm{M}}| = (\sqrt{3}/2)L_{\mathrm{M}}^{2}$.

\subsection{Effective Continuum Model}
\label{sec:effective_continuum_model}

For the situation where the moir\'{e} lattice constant is much larger than the graphene lattice constant, the coupling between the two layers can be described by a low energy effective continuum model \cite{LopesDosSantos2007,Bistritzer2011,Kindermann2011,LopesdosSantos2012,Moon2013,Koshino2015,Koshino2015b,Koshino2018}. In such systems, the separation between the Dirac points at different valleys is sufficiently large that inter-valley mixing can be safely neglected and the full Hamiltonian separates into two independent Hamiltonians, each describing the electronic properties of a single valley. These Hamiltonians read \cite{Koshino2019,Chebrolu2019}
\begin{equation}
H^{(\xi)}_{\mathrm{AB-AB}} = \left(\begin{array}{cccc}
H_{1}^{(\xi)}(\mathbf{k}_{1}) & g^{\dagger}(\mathbf{k}_{1}) & 0 & 0\\
g(\mathbf{k}_{1}) & \tilde{H}_{1}^{(\xi)}(\mathbf{k}_{1}) & U^{\dagger} & 0\\
0 & U & H_{2}^{(\xi)}(\mathbf{k}_{2}) & g^{\dagger}(\mathbf{k}_{2})\\
0 & 0 & g(\mathbf{k}_{2}) & \tilde{H}_{2}^{(\xi)}(\mathbf{k}_{2})
\end{array}\right),
\label{eq_H_ABAB}
\end{equation}
for AB-AB stacked TDBG and
\begin{equation}
H^{(\xi)}_{\mathrm{AB-BA}} = \left(\begin{array}{cccc}
H_{1}^{(\xi)}(\mathbf{k}_{1}) & g^{\dagger}(\mathbf{k}_{1}) & 0 & 0\\
g(\mathbf{k}_{1}) & \tilde{H}_{1}^{(\xi)}(\mathbf{k}_{1}) & U^{\dagger} & 0\\
0 & U & \tilde{H}_{2}^{(\xi)}(\mathbf{k}_{2}) & g(\mathbf{k}_{2})\\
0 & 0 & g^{\dagger}(\mathbf{k}_{2}) & H_{2}^{(\xi)}(\mathbf{k}_{2})
\end{array}\right),
\label{eq_H_ABBA}
\end{equation}
for AB-BA stacked TDBG where $\textbf{k}_l = R(\pm\theta/2)(\textbf{k}-\textbf{K}_\xi^{(l)})$
with $\pm$ for layer index $l$ of 1 and 2, respectively, and
\begin{gather}
H_{l}^{(\xi)}(\mathbf{k}_l) = \left(\begin{array}{cc}
0 & -\hbar v_{F}k^{(\xi,l)}_{-}\\
-\hbar v_{F}k^{(\xi,l)}_{+} & \Delta
\end{array}\right),\\
\tilde{H}_{l}^{(\xi)}(\mathbf{k}_l) = \left(\begin{array}{cc}
\Delta & -\hbar v_{F}k^{(\xi,l)}_{-}\\
-\hbar v_{F}k^{(\xi,l)}_{+} & 0
\end{array}\right),
\end{gather}
is the Hamiltonian for monolayer graphene and
\begin{equation}
g(\mathbf{k}_l) = \left(\begin{array}{cc}
\hbar v_{4}k^{(\xi,l)}_{+} & \gamma_{1}\\
\hbar v_{3}k^{(\xi,l)}_{-} & \hbar v_{4}k^{(\xi,l)}_{+}
\end{array}\right),
\end{equation}
is the intra-bilayer coupling. 
Here, $v_{F}$ ($\approx 0.8\times10^6\,\mathrm{m/s}$) is a Fermi velocity,
$k^{(\xi,l)}_{\pm} = e^{\pm i\xi\eta^{(l)}}(\xi k_{x}\pm ik_{y})$,
where $\eta^{(l)}$ is the angle between $\mathbf{a}_{1}^{(l)}$ and the $x$-axis (which in this case is $\eta^{(1)/(2)} = \mp \theta/2$),
$\Delta = 0.05\,\mathrm{eV}$ is the on-site potential of the lattice sites
that are vertically aligned with the lattice sites in the adjacent layer of the bilayer,
$\gamma_{1} = 0.4\,\mathrm{eV}$ is the interaction strength between the $A$ and $B$ sites in the upper and lower layers of the graphene bilayer,
$v_3 = 1.036\times 10^5\,\mathrm{m/s}$ and $v_4 = 0.143\times 10^4\,\mathrm{m/s}$
are responsible for the trigonal warping and electron-hole asymmetry, respectively \cite{McCann2013}.
Finally, the matrix
\begin{multline}
U = \left(\begin{array}{cc}
u & u'\\
u' & u
\end{array}\right) +
\left(\begin{array}{cc}
u & u'\omega^{-\xi}\\
u'\omega^{\xi} & u
\end{array}\right)e^{i\xi\mathbf{G}_{1}^{\mathrm{M}}\cdot\mathbf{r}}\\
+ \left(\begin{array}{cc}
u & u'\omega^{\xi}\\
u'\omega^{-\xi} & u
\end{array}\right)e^{i\xi(\mathbf{G}_{1}^{\mathrm{M}}+\mathbf{G}_{2}^{\mathrm{M}})\cdot\mathbf{r}},
\end{multline}
is the inter-bilayer coupling Hamiltonian with $\omega = e^{2\pi i/3}$ and $u=0.07974\,\mathrm{eV}$ and $u'=0.09754\,\mathrm{eV}$ coupling constants that give the strength of the interaction between like ($A \leftrightarrow A$, $B \leftrightarrow B$) and opposing ($A \leftrightarrow B$) sublattices, respectively, between the lower layer of the upper bilayer and the upper layer of the lower bilayer \cite{Koshino2018}.

The addition of an electrostatic potential can be modeled by the addition of a diagonal matrix to the Hamiltonian $H\rightarrow H+V_{b}$ where
\begin{equation}
V_{b} = \left(\begin{array}{cccc}
\frac{3V}{2}\mathbb{I} & 0 & 0 & 0\\
0 & \frac{V}{2}\mathbb{I} & 0 & 0\\
0 & 0 & -\frac{V}{2}\mathbb{I} & 0\\
0 & 0 & 0 & -\frac{3V}{2}\mathbb{I}
\end{array}\right),
\label{V}
\end{equation}
with $V$ the applied electrostatic potential and $\mathbb{I}$ the $2 \times 2$ unit matrix.

\subsection{Electronic Structure in a Magnetic Field}
\label{sec:electronic_structure_in_a_magnetic_field}

Consider TDBG subject to a uniform perpendicular magnetic field $\mathbf{B} = \bm{\nabla}\times\mathbf{A} = (0,0,B)$. In the following we will neglect the Zeeman effect. In general, when a material is subjected to a magnetic field, the periodicity of the lattice is lost owing to the spatial dependence of the vector potential, which, in the Landau gauge, reads $\mathbf{A} = (0,Bx,0)$. However, for certain values of the magnetic field - specifically, when the number of quanta of magnetic flux per unit cell is a rational number (i.e. $\Phi/\Phi_{0} = p/q$ where $p$ and $q$ are co-prime integers, $\Phi = BA$ is the flux through the unit cell and $\Phi_{0} = h/e$ is the quantum of magnetic flux) then, in the Landau gauge with the y-axis of the coordinate system parallel to $\mathbf{L}_{1}$, one can introduce a periodic magnetic unit cell with lattice vectors $\tilde{\mathbf{L}}_{1} = \mathbf{L}_{1}$ and $\tilde{\mathbf{L}}_{2} = q\mathbf{L}_{2}$ \cite{brown1969aspects,Xiao2010}. Hence, one can find `magnetic' Bloch conditions for this enlarged unit cell
\begin{gather}
\Psi_{\mathbf{k}}(\mathbf{r}+\tilde{\mathbf{L}}_{1}) = e^{i\mathbf{k}\cdot\tilde{\mathbf{L}}_{1}}\Psi_{\mathbf{k}}(\mathbf{r}).\label{mbc2} \\
\Psi_{\mathbf{k}}(\mathbf{r}+\tilde{\mathbf{L}}_{2}) = e^{i\mathbf{k}\cdot\tilde{\mathbf{L}}_{2}}e^{-i(e/\hbar)(\mathbf{A}-\mathbf{B}\times\mathbf{r})\cdot\tilde{\mathbf{L}}_{2}}\Psi_{\mathbf{k}}(\mathbf{r}),\label{mbc1}
\end{gather}
Within each layer, one can construct a wave function that obeys the magnetic Bloch conditions in Eqs.\,\eqref{mbc2} and \eqref{mbc1} from the Landau levels of monolayer graphene. The effective continuum Landau levels for the $K$ ($\xi=+1$) and $K'$ ($\xi=-1$) valleys in monolayer graphene read \cite{Ando2005, Shon1998}
\begin{gather}
\Psi^{(l)}_{n,k_{y},+}(\mathbf{r}) = C_{n}e^{ik_{y}y}\left(\begin{array}{c}
-i\mathrm{sgn}(n)\phi_{|n|-1,k_{y}}(x)\\
-e^{i\eta^{(l)}}\phi_{|n|,k_{y}}(x)
\end{array}\right)e^{i\mathbf{K}^{(l)}_{+}\cdot\mathbf{r}},\\
\Psi^{(l)}_{n,k_{y},-}(\mathbf{r}) = C_{n}e^{ik_{y}y}\left(\begin{array}{c}
e^{i\eta^{(l)}}\phi_{|n|,k_{y}}(x)\\
-i\mathrm{sgn}(n)\phi_{|n|-1,k_{y}}(x)
\end{array}\right)e^{i\mathbf{K}^{(l)}_{-}\cdot\mathbf{r}},
\end{gather}
respectively, with the upper and lower components of the vector referring to the $A$ and $B$ sublattices respectively. Here, $n$ is the Landau level index, $k_y$ is the wave vector in the $y$-direction and $\eta^{(l)}$ is, again, the angle between $\mathbf{a}_{1}^{(l)}$ and the $x$-axis. The single particle Landau level is defined in terms of the Hermite polynomial, $H_{n}(z)$, as $\phi_{n,k_{y}}(x) = (2^{n}n!\sqrt{\pi}l_{B})^{-1/2}e^{-z^2/2}H_{n}(z)$ with $z=(x+k_{y}l_{B}^{2})/l_{B}$ and $l_{B} = \sqrt{\hbar/(eB)}$ \cite{Shon1998, Zheng2002}. The normalization coefficient reads $C_{n} = 1$ for $n=0$ and $C_{n} = 1/\sqrt{2}$ for $n\neq 0$. To find a wave function that satisfies Eqs.\,\eqref{mbc2} and \eqref{mbc1} one needs to combine Landau levels at different $k_{y}$ via
\begin{equation}
\Psi^{(l)}_{n,\textbf{k},m,\xi} = \sum_{j=-\infty}^{\infty}\alpha^{j} \Psi^{(l)}_{n,k_{y}^m,\xi}(\mathbf{r}),
\label{wvfn}
\end{equation}
with
\begin{gather}
\alpha = e^{i(\mathbf{k}-\mathbf{K}_{\xi}^{(l)})\cdot(\tilde{\mathbf{L}}_{2}-q\tilde{\mathbf{L}}_{1}/2)} e^{i \pi p q (j+1)/2} e^{i \pi q m},\\
k_{y}^{m} = k_{y} - (\mathbf{K}_{\xi}^{(l)})_{y}-\frac{2\pi}{L_M}(pj+m),
\end{gather}
and the $m$ index running from $0$ to $p-1$.

The electronic properties of TDBG in a magnetic field can be found by constructing a matrix Hamiltonian for each valley in the basis of the wave functions given in Eq.\,\eqref{wvfn}
\begin{equation}
H_{n,n',m,m',l,l'}^{(\xi)}(\mathbf{k}) = \langle\Psi^{(l')}_{n',k_{y}^{m'},\xi}|H^{(\xi)}_{i}|\Psi^{(l)}_{n,k_{y}^{m},\xi}\rangle,
\end{equation}
with $i$ indicating AB-AB or AB-BA stacking. The diagonal elements of the matrix Hamiltonian reduces to 
\begin{equation}
H_{n,n',m,m',l,l}^{(\xi)}(\mathbf{k}) = \varepsilon_{n}\delta_{n,n'}\delta_{m,m'},
\end{equation}
where $\varepsilon_{n} = \hbar\omega_{B}\mathrm{sgn}(n)\sqrt{|n|}$ is the single particle Landau level energy with $\omega_{B} = \sqrt{2v_{F}^2eB/\hbar}$ \cite{Shon1998, Zheng2002}. The intra-bilayer matrix elements can be evaluated by noting that the momentum operators $\hbar k_\pm^\xi$ become canonical momentum $\GVec{\pi}_\pm^\xi=\hbar k_\pm^\xi + e\Vec{A}$ in magnetic fields as well as the relation $\GVec{\pi}_+^{\xi=+1}=(\sqrt{2}\hbar/l_B)a^\dagger$, $\GVec{\pi}_-^{\xi=+1}=(\sqrt{2}\hbar/l_B)a$, $\GVec{\pi}_+^{\xi=-1}=-(\sqrt{2}\hbar/l_B)a$, $\GVec{\pi}_-^{\xi=-1}=-(\sqrt{2}\hbar/l_B)a^\dagger$, where $a^\dagger$ and $a$ are raising and lowering operators on the Landau levels. The inter-bilayer matrix elements can be evaluated using the identity \cite{Pfannkuche1992}
\begin{align}
&\langle\phi_{n',k_{y}'}(x)e^{ik_{y}'y}|e^{i\mathbf{G}\cdot\mathbf{r}}|e^{ik_{y}y}\phi_{n,k_{y}}(x)\rangle,\nonumber\\
&\quad = \delta_{k_{y}',k_{y}+G_{y}}\sqrt{\frac{\lambda!}{\Lambda!}}\left(\frac{G_{x} + iG_{y}}{|\mathbf{G}|}\right)^{n-n'}\left(\frac{i|\mathbf{G}|l_{B}}{\sqrt{2}}\right)^{|n-n'|}\nonumber\\
&\qquad \times e^{-|\mathbf{G}|^{2}l_{b}^{2}/4}e^{-il_{B}^{2}G_{x}(k_{y}' + k_{y})/2}L_{\lambda}^{|n-n'|}\left(\frac{|\mathbf{G}|^2l_{B}^{2}}{2}\right),
\end{align}
where $\lambda = \mathrm{min}(n,n')$, $\Lambda = \mathrm{max}(n,n')$ and $L^{\alpha}_{\lambda}(x)$ is an associated Laguerre polynomial. This matrix Hamiltonian is unbounded in both $n$ and $j$. However, by applying magnetic Bloch conditions one can see that the state with $j=1$ and $m=0$ is equivalent to the state with $j=0$ and $m=p$ which is just the state $j=0$ and $m=0$ with the addition of a phase. Thus, one only needs to consider a single cycle of $m \in [0, p-1]$ with the appropriate periodic boundary conditions. The $n$ index relates to the energy of the Landau level basis but one can truncate the Hamiltonian at an energy at which the Landau level only weakly affect the low energy spectrum. This cutoff energy must be significantly larger that the interlayer coupling characterized by the coupling constants $u$ and $u'$. This bounded matrix can then be diagonalized to find the energy spectrum of TDBG.

The band structures of TDBG can be probed experimentally by measuring the Hall current. When the Fermi energy, $\varepsilon_{F}$, lies within a band gap, the normalized electron density can be found by summing number of bands between the Fermi energy and the charge neutrality point ($n=0$). This normalized electron density, $n/n_0$, with $n_0 = 1/A$ is the electron density per Bloch band and can be related to the number of flux quanta per unit cell, $\Phi/\Phi_{0}$ via the relation \cite{thouless1982quantized,Kohmoto1985}
\begin{equation}
\frac{n}{n_0} = t\frac{\Phi}{\Phi_{0}} + s,
\label{tknn}
\end{equation}
where $t$ and $s$ are topologically invariant integers
which represent the quantized Hall conductivity $\sigma_{xy} = -te^2/h$ and the Bloch band filling at each gap, respectively.
By differentiating Eq.\,\eqref{tknn} with respect to the magnetic field one arrives at the Streda-Widom formula \cite{Streda1982,widom1982thermodynamic}
\begin{equation}
\sigma_{xy} = -e\left(\frac{\partial n}{\partial B}\right)_{\varepsilon_{F}},
\end{equation}
from which one can compute the Hall conductivity $\sigma_{xy} = -te^2/h$ directly.

\section{Results and Discussion}
\label{sec:results_and_discussion}

\subsection{Symmetry argument}
\label{sec:symmetry_argument}

Before we calculate the energy spectrum,
we perform a symmetry analysis on the lattice structure and Hamiltonian matrix
to explain the features (such as the valley degeneracy and electron-hole symmetry) of the energy spectrum.
The lattice structure of the AB-AB stacked TDBG has $C_{2x}$ symmetry \cite{Koshino2019},
which makes $E_\alpha^{(\xi)}$ under $+B$ identical to $E_\alpha^{(\xi)}$ under $-B$.
Since the time reversal symmetry $T$
requires $E_\alpha^{(\xi)}$ under $+B$ identical to $E_\alpha^{(-\xi)}$ under $-B$,
the combined operation $C_{2x} T$ guarantees that
$E_{\alpha}^{(\xi)}$ under $+B$ is identical to
$E_{\alpha}^{(-\xi)}$ under $+B$, i.e.,
the energy spectrum of AB-AB stacked TDBLG is valley degenerate. 
In the presence of interlayer potential asymmetry $V_b$ [Eq.\,\eqref{V}], however,
the valley degeneracy is lifted
since $V_b$ breaks the $C_{2x}$ lattice symmetry.
On the contrary, 
the AB-BA stacked TDBG does not have the valley degeneracy,
since its lattice symmetry, $C_{2y}$,
makes $E_\alpha^{(\xi)}$ under $+B$ identical to $E_\alpha^{(-\xi)}$ under $-B$,
which is identical to the relation granted by the time reversal symmetry.
Thus, AB-BA stacking has no symmetry protecting the degeneracy between opposite valleys at the same $B$.
The same conclusions can also be derived in terms of the wave functions of Landau levels
(see Appendix \ref{appendix1}).

We have another type of symmetry between the electron side and the hole side of the spectrum.
	Although the electron-hole symmetry is not strictly obeyed in the real system, it provides useful insight on some of the properties of similar materials.
	Such approximate electron-hole symmetry is rigorous in the "minimal model",
	which neglects the relatively small parameters $v_3$, $v_4$, $\Delta$,
	and the rotation matrix $R(\pm\theta/2)$ in the definition of $\textbf{k}_1$ and $\textbf{k}_2$.
	The minimal AB-AB Hamiltonian exhibits a fictitious particle-hole symmetry
	\begin{align}
	\Sigma^{-1} H_{\mathrm{AB}-\mathrm{AB}}^{(\xi)} \Sigma = -H_{\mathrm{AB-AB}}^{(-\xi)}, \nonumber\\
	\Sigma=\begin{pmatrix}
	&&&\sigma_{x} \\
	&&-\sigma_{x}& \\
	&\sigma_{x} && \\
	-\sigma_{x} &&&
	\end{pmatrix},
	\label{eq_H_ABAB_sym}
	\end{align}
	similar to that of the AB-AB stacked TDBG in the absence of magnetic fields \cite{Koshino2019},
	but now for opposite monolayer valleys.
	This symmetry leads to the electron-hole symmetry between different valleys
	\begin{equation}
	E_{n,\textbf{k}}^{(\xi)} = -E_{-n,\textbf{k}}^{(-\xi)},
	\label{eq_E_sym_ABAB}
	\end{equation}
	where $n$ is the band index and $\Vec{k}$ is the magnetic Bloch wavenumber.
	Note that Eqs.\,\eqref{eq_H_ABAB_sym} and \eqref{eq_E_sym_ABAB} hold
	regardless of whether magnetic fields $B$ or interlayer potential asymmetry $V_b$ exist.
	Equation \eqref{eq_E_sym_ABAB}, together with the valley degeneracy from the combination of the $C_{2x}$ lattice symmetry
	and the time reversal symmetry in the absence of $V_b$,
	leads to the electron-hole symmetry $E_{n,\textbf{k}}^{(\xi)}=-E_{-n,\textbf{k}}^{(\xi)}$ within the same valley in the AB-AB stacking.
	In the presence of $V_b$, the electron-hole symmetry is broken
	according to the lift of the $C_{2x}$ symmetry.

	The AB-BA Hamiltonian, on the contrary,
	satisfies a different type of symmetry \cite{Koshino2019}

\begin{align}
(\Sigma'^{-1} \tilde{P}) H_{\mathrm{AB-BA}}^{(\xi)} (\tilde{P} \Sigma') = -H_{\mathrm{AB-BA}}^{(\xi)}, \nonumber\\
\Sigma'=\begin{pmatrix}
&&&\mathbb{I} \\
&&-\mathbb{I}& \\
&\mathbb{I} && \\
-\mathbb{I} &&&
\end{pmatrix},
\label{eq_H_ABBA_sym}
\end{align}
	within the same monolayer valley,
	where $\tilde{P}$ is a space inversion operator which works on the envelope function as $\tilde{P} F_X(\textbf{r}) = F_X(-\textbf{r})$,
	while it does not change the sublattice degree of freedom $(X=A_1, B_1, ...)$.
	This symmetry leads to the electron-hole symmetry within the same valley
	\begin{equation}
	E_{n,\textbf{k}}^{(\xi)} = -E_{-n,-\textbf{k}}^{(\xi)}.
	\label{eq_E_sym_ABBA}
	\end{equation}
	Note that Eqs.\,\eqref{eq_H_ABBA_sym} and \eqref{eq_E_sym_ABBA} hold
	regardless of whether $B$ or $V_b$ exist.


	If we consider all the band parameters in the Hamiltonian (hereafter "full model"),
	Eqs.\,\eqref{eq_H_ABAB_sym} and \eqref{eq_H_ABBA_sym} are no longer valid,
	and nor are Eqs.\,\eqref{eq_E_sym_ABAB} and \eqref{eq_E_sym_ABBA}.
	However, the overall spectrum,
	e.g., the distribution of the major gaps,
	is qualitatively consistent with that in the minimal model,
	since the parameters neglected in the minimal model
	are relatively small.

\subsection{Energy spectrum}
\label{sec:energy_spectrum}

\begin{figure*}
\centering
\includegraphics[width=1.9\columnwidth]{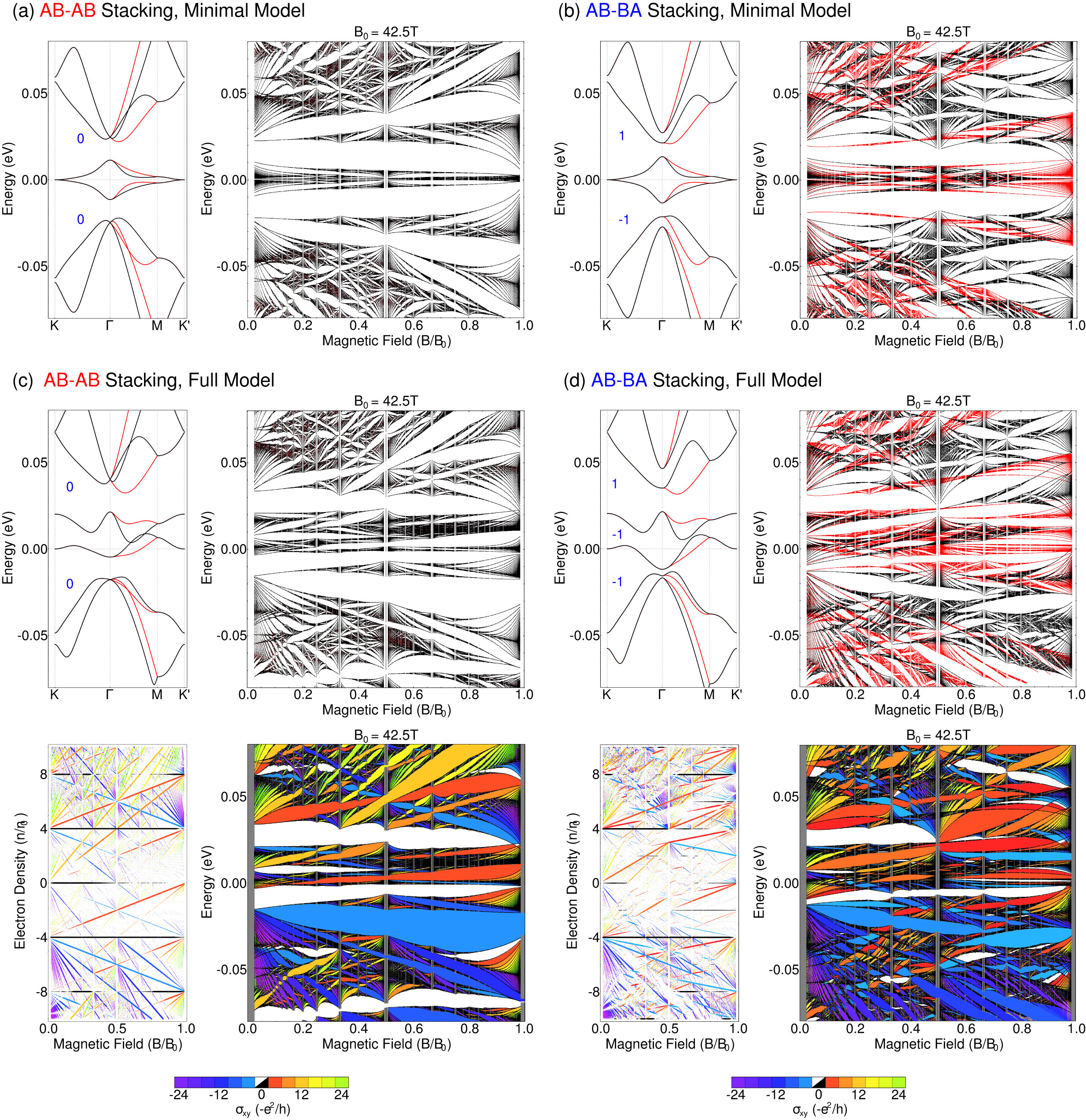}
\caption{(Colour online) (a) and (b): Band structures at $B=0\,\mathrm{T}$ (left panel) and energy spectrum at $B\ne0\,\mathrm{T}$ plotted against $B/B_0$ (=$\Phi/\Phi_0$, $B_0 \equiv \Phi_0/A$) (right panel) for TDBG for (a) AB-AB stacking and (b) AB-BA stacking in a minimal model. The black and red bands indicate those originating from the $K$ and $K'$ valley, respectively. Numbers in the band structures at $B=0\,\mathrm{T}$ indicate the integrated Chern numbers summed over all the energy bands of $K$ bands, while the numbers of $K'$ bands (not shown) are opposite in sign. (c) and (d): Upper panels show the plots similar to (a) and (b) for TDBG at (c) AB-AB stacking and (d) AB-BA stacking in a full model. The lower right panel shows the plot similar to the upper right panel with each gap filled with shading corresponds to the quantized Hall conductivity $\sigma_{xy}$ in units of $-e^2/h$. The lower left panel is the Wannier diagram, which shows the position of energy gaps in the space of charge density and magnetic field. Each gap is plotted as a point of which radius is proportional to the gap width and the color represents $\sigma_{xy}$.
The two lower panels share the same color map for the values of $\sigma_{xy}$, except for $\sigma_{xy}=0$ which is plotted in white and black in the left and right panels, respectively.}
\label{fig2}
\end{figure*}

We compute the band structures for TDBG under a perpendicular magnetic field in a minimal model and also in a full model.
In all cases, we use the twist angle $\theta$ of $1.33^\circ$, close to the angles at which superconductivity was observed in the experiments \cite{shen2019observation,liu2019spin}.

The first two panels in Figs.\,\ref{fig2}(a) and (c)
show the band structures at $B=0\,\mathrm{T}$ and
energy spectrum in magnetic field of AB-AB stacked TDBG calculated by the minimal model and full model, respectively.
The black and red bands indicate those originating from the $K$ and $K'$ valley, respectively,
and the numbers in the band structures at $B=0\,\mathrm{T}$ indicate
the integrated Chern numbers summed over all the occupied bands of K valley below the energy gap.
The lower right panel in Fig.\,\ref{fig2}(c) show the energy spectrum
with each gap filled with shading corresponds to the quantized Hall conductivity $\sigma_{xy}$.
And the lower left panel in Fig.\,\ref{fig2}(c) is the Wannier diagram \cite{Wannier1978},
which shows all the spectral gaps plotted against the normalized electron density $n/n_0$
and the number of magnetic flux quanta per unit cell $\Phi/\Phi_0 (=B/B_0)$;
we plot each point in colour and size according to $\sigma_{xy}$ and the size of the gap, respectively.
According to Eq.\,\eqref{tknn}, all the gaps are constrained to linear trajectories
starting from $s \in \mathbb{Z}$ at $B\rightarrow0\,\mathrm{T}$.
From the slope and y-intercept of these lines,
we can get the topological invariant $t$ and $s$, respectively,
where $\sigma_{xy} = -t e^2/h$ and $s=\lim_{B\rightarrow 0} (n/n_0)$ for the electron density $n$ below the gap.
Thus, changing $\sigma_{xy}$ constitutes a topological phase transition where the values of the topological invariants $t$ and $s$ are changed. 
Figures \ref{fig2}(b) and (d) show the plots similar to Figs.\,\ref{fig2}(a) and (c), but for the AB-BA stacked TDBG.

Although the AB-AB stacked TDBG and AB-BA stacked TDBG exhibit
almost the same $B=0\,\mathrm{T}$ band structures,
their energy spectrum in magnetic fields are totally different;
the energy spectrum from monolayer $K$ and $K'$ valleys are degenerate in AB-AB stacking,
while the degeneracy is lifted in AB-BA stacking.
As explained in Sec.\,\ref{sec:symmetry_argument},
the difference arises because
the $C_{2x}$ lattice symmetry of AB-AB stacking, together with the time reversal symmetry protects the valley degeneracy,
while the $C_{2y}$ symmetry of AB-BA stacking cannot.
In the minimal model, in addition,
the energy spectra of the AB-AB stacking and AB-BA stacking exhibit the electron-hole symmetry
between the opposite [Eq.\,\eqref{eq_E_sym_ABAB}] and the same [Eq.\,\eqref{eq_E_sym_ABBA}] valley, respectively.
While such symmetry is absence in the spectrum in the full model,
since the extra coupling terms lead to a broadening of the central band at $E=0$
as well as the hybridization of Landau levels with their indices differ by 3,
the overall spectrum is qualitatively consistent with that in the minimal model,
except the gap at the charge neutrality point.
Moreover, those coupling terms do not change the spatial symmetry of TDBG,
thus, the AB-AB stacked TDBG in the full model still keeps the valley degeneracy,
while the AB-BA stacked TDBG lacks the degeneracy.

In weak magnetic fields of $B/B_0 < 0.1$,
the lifting of the valley degeneracy in the AB-BA spectrum
is observed as a split of Landau levels.
As a result, the degeneracy of the Landau levels of AB-BA is
half of the Landau levels of AB-AB where the valleys are completely degenerate.
Generally we have 4-fold degeneracy from spin and valley in AB-AB, while 2-fold degeneracy in AB-BA.
Around the charge neutrality point, however,
the Landau levels in the full model become 12-fold degenerate in AB-AB and 6-fold degenerate in AB-BA.
The extra factor of 3
comes from the trigonal-warping of the $B=0\,\mathrm{T}$ band structures;
as Figs.\,\ref{fig2}(c) and (d) show,
the full model introduces three band minima for each spin-valley sector
at the wave vector between $\bar{\Gamma}$ and $\bar{M}$.
Thus, we have the Hall conductivity sequence $0, \pm 12, \pm 24 …$ in units of $-e^2/h$ for AB-AB,
while we have $0, \pm 6, \pm 12 …$ in AB-BA.
Note that, in minimal model [Figs.\,\ref{fig2}(a) and (b)],
the Landau level degeneracy around the charge neutral point 
are 8-fold in AB-AB and 4-fold in AB-BA,
because the trigonal warping is absent and the Landau levels originate from the degenerate $\bar{K}$ and $\bar{K}'$.



In large magnetic fields of $B/B_0 \ge 0.1$,
each Landau level in both the AB-AB and AB-BA configurations
evolves to a complicated fractal energy spectrum, aka Hofstadter butterfly \cite{Hofstadter1976},
due to the competition between the periodic potential and magnetic field.
In AB-BA stacking, the $K$ and $K'$ spectrum become completely un-correlated.
In certain energy regions, as a result, we have many inter-valley Landau level crossings,
while at other regions we have completely valley-polarized states.
The spectrum in AB-AB stacking, on the contrary,
still keeps the valley degeneracy
from the $C_{2x}$ lattice symmetry.
As a result, the spectrum
exhibits many gaps surviving over a wide range of magnetic fields
due to the lack of inter-valley Landau level crossing.


The valley degeneracy of Landau level is also correlated to the valley Hall conductivity at $B=0\,\mathrm{T}$.
In Figs.\,\ref{fig2}(a) and (b), the numbers in the $B=0\,\mathrm{T}$ band structures 
indicate the single-valley Chern numbers 
\begin{equation}
C=\frac{1}{2\pi} \int_{\mathrm{MBZ}} \mathcal{F}_{\textbf{k}} d\textbf{k}
\label{eq_chern_number}
\end{equation}
defined for every energy gap of $K$ band \cite{Koshino2019}.
The corresponding Chern number for $K'$ valley is given by $-C$.
The single-valley Hall conductivity is given by $\sigma^{(\xi)}_{xy} = \xi (e^2/h) C$ for valley $\xi = \pm$.
As argued in the beginning of this section, the Hall conductivity coincides with the gradient in the Wannier diagram.
As magnetic field strength increases, the number of states below a gap increases (decreases) when $\sigma_{xy}$ is negative (positive).
The same is true also for the single valley Hall conductivity.
In AB-BA stacking [Fig. 2(d)], for instance, 
the first gap in the conduction band has $C=1$ (i.e., $\sigma^{(+)}_{xy}>0$ and $\sigma^{(-)}_{xy}<0$), and therefore the number 
of states of $\xi=+$ ($-$) valley below the gap decreases (increases) in increasing $B$ (see Appendix \ref{appendix2} and Fig.\,\ref{fig_s1} for more detail).
As a result, the energy gaps of the opposite valleys trace opposite slopes in the Wannier diagram,
and finally they lose the overlap at $B/B_0 \approx 0.5$, closing the gap in the whole spectrum [Fig.\,\ref{fig2}(d)].
In AB-AB stacking, on the other hand,
the valley Chern number is all zero from the symmetry \cite{Koshino2019}.
Therefore, the spectrum from both valleys are always degenerate and less likely to be masked in increasing B-field,
because the gaps of opposite valleys have the same slope in the Wannier diagram and never lose the overlap.
Indeed, the first gap in the conduction band remains open up to $B/B_0 \approx 1$
[Fig.\,\ref{fig2}(c)].

\begin{figure*}
\centering
\includegraphics[width=1.9\columnwidth]{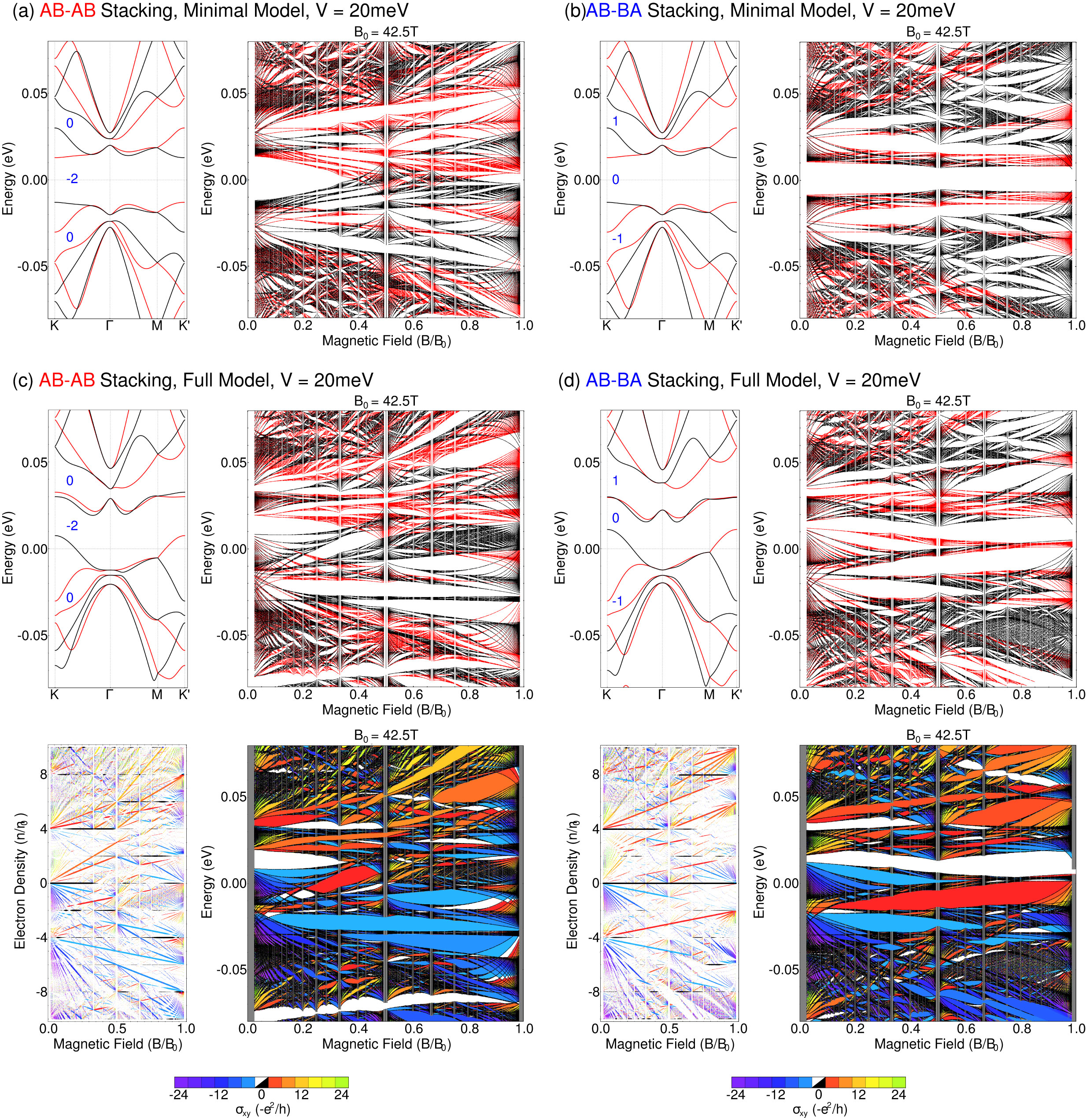}
\caption{(Colour online) Plots similar to Fig.\,\ref{fig2} for TDBG with an electrostatic potential Eq.\,\eqref{V} with $V=20\,\mathrm{meV}$.}
\label{fig3}
\end{figure*}

Figure \ref{fig3} shows the plot similar to Fig.\,\ref{fig2} for TDBGs subject to the interlayer potential asymmetry $V_b$ [Eq.\,\eqref{V}] with $V = 20\,\mathrm{meV}$.
A notable feature is that now the valley degeneracy in the AB-AB stacked TDBG is lifted,
since the interlayer potential asymmetry breaks the $C_{2x}$ symmetry of the system.

In the minimal model,
the energy spectrum of the $K$ valley of the AB-AB stacked TDBG
can be obtained by inverting the sign of the energy spectrum of the $K'$ valley.
In the AB-BA stacked TDBG, on the contrary, the energy spectrum of each valley
is symmetric with respect to the charge neutrality point. 
This is, again, a result of the electron-hole symmetry
between the opposite valleys in the AB-AB stacking [Eq.\,\eqref{eq_E_sym_ABAB}]
and between the same valleys in the AB-BA stacking [Eq.\,\eqref{eq_E_sym_ABBA}], respectively,
which is valid even in the presence of $V_b$.
In the full model, 
although such electron-hole symmetries has already been broken in the Hamiltonian in the absence of $V_b$,
the overall energy spectrum is qualitatively consistent with that in the minimal model.

The electron-hole symmetry is also correlated to the valley Hall conductivity.
In both the minimal model and full model,
the gap at the charge neutrality point in the AB-AB stacked TDBG has
a valley Chern number $\mp2$ for $\xi=\pm$ valley.
Accordingly, the gap is closed as magnetic field increases,
since the gap in the $K$ and $K'$ valleys moves to positive and negative electron densities, respectively.
On the contrary, the gap in the AB-BA stacked TDBG remains open,
since the Chern number is zero at the gap.

\section{Conclusion}
\label{sec:conclusion}

We have investigated the energy spectrum and Hall conductivities of the twisted double bilayer graphene in uniform magnetic field.
We found that 
the two different configurations, AB-AB and AB-BA stacking, which have different valley Chern numbers,
exhibit totally different energy spectrum
in spite of the similarity between their band structures in the absence of magnetic fields;
The energy spectrum in the AB-AB stacking is valley degenerate,
while the degeneracy can be lifted by interlayer potential asymmetry.
On the contrary, the energy spectrum in the AB-BA has no valley degeneracy in any case.
From the perspective of the lattice symmetry,
the valley degeneracy in AB-AB is protected by the combination of the $C_{2x}$ lattice symmetry and the time reversal symmetry,
which can be lifted by interlayer potential asymmetry,
while AB-BA has no such symmetry.
From the perspective of the band topology,
the valley Hall conductivity vanishes at all the gaps in AB-AB,
hence the spectrum is valley degenerated.
On the other hand,
the gaps in AB-BA have non-zero valley Hall conductivities,
hence the spectrum in opposite valleys evolve in different manners.
Therefore, in certain energy regions in AB-BA,
we have many inter-valley Landau level crossings, while at other regions we have complete valley-polarized states.
We also found the fictitious electron-hole symmetry in both systems,
which is weakly obeyed even if we consider all the parameters and also in the presence of interlayer potential asymmetry.


\section{Acknowledgments}

J.A.C was supported by the National Science Foundation of China Research Grant No. 11750110420.
M.K. was supported by JSPS KAKENHI Grants No. JP17K05496, No. JP20H01840, and No. JP16K17755.
P.M. acknowledges the support by Science and Technology Commission of Shanghai Municipality (Shanghai Natural Science Grants,
grant no. 19ZR1436400).
J.A.C. and P.M. were supported by
the NYU-ECNU Institute of Physics at NYU Shanghai.
This research was carried out on the High Performance Computing resources at NYU Shanghai.


\appendix

\section{Valley degeneracy in TDBG}
\label{appendix1}
Besides the symmetry analysis in Sec.\,\ref{sec:symmetry_argument},
the valley degeneracy in the AB-AB stacking
as well as the absence of such degeneracy in the AB-BA stacking
can also be explained in terms of the wave functions of Landau levels.
For AB-AB stacking, the Landau levels in bilayers 1 and 2 are written in the form of
$\Psi^{(1)}_{n,\xi}(\mathbf{r}) = (c_1 \phi_{n-\xi},c_2 \phi_{n},c_3 \phi_{n},c_4 \phi_{n+\xi})$
and
$\Psi^{(2)}_{n,\xi}(\mathbf{r}) = (c_5 \phi_{n-\xi},c_6 \phi_{n},c_7 \phi_{n},c_8 \phi_{n+\xi})$,
respectively \cite{Pereira2007, Rozhkov2016}.
The inter-bilayer coupling couples the lower two components of $\Psi^{(1)}_{n,\xi}$ to the upper two components of $\Psi^{(2)}_{n,\xi}$
and, hence, Landau levels that are coupled are $(\phi_{n}, \phi_{n+1})$ and $(\phi_{n-1}, \phi_{n})$ for the $K$ valley, and $(\phi_{n}, \phi_{n-1})$ and $(\phi_{n+1}, \phi_{n})$ for the $K'$ valley. As the intra-bilayer coupling Hamiltonian is a real, symmetric matrix, it leads to the same coupling in both the valleys and, hence, the same band structures.
For AB-BA stacking, on the other hand, the Landau levels in bilayers 1 and 2 are written as
$\Psi^{(1)}_{n,\xi}(\mathbf{r}) = (c_1 \phi_{n-\xi},c_2 \phi_{n},c_3 \phi_{n},c_4 \phi_{n+\xi})$
and
$\Psi^{(2)}_{n,\xi}(\mathbf{r}) = (c_5 \phi_{n},c_6 \phi_{n+\xi},c_7 \phi_{n-\xi},c_8 \phi_{n})$,
respectively.
Thus, the Landau levels that are coupled by the inter-bilayer coupling are $(\phi_{n}, \phi_{n+1})$ and $(\phi_{n}, \phi_{n+1})$ in the $K$ valley and $(\phi_{n}, \phi_{n-1})$ and $(\phi_{n}, \phi_{n-1})$ in the $K'$ valley. Thus for AB-BA stacking the inter-bilayer coupling is different in the two valleys and, hence, the valley degeneracy is lifted.

\section{Valley Hall conductivity in the minimal model}
\label{appendix2}

\begin{figure*}[t]
	\centering
	\includegraphics[width=1.9\columnwidth]{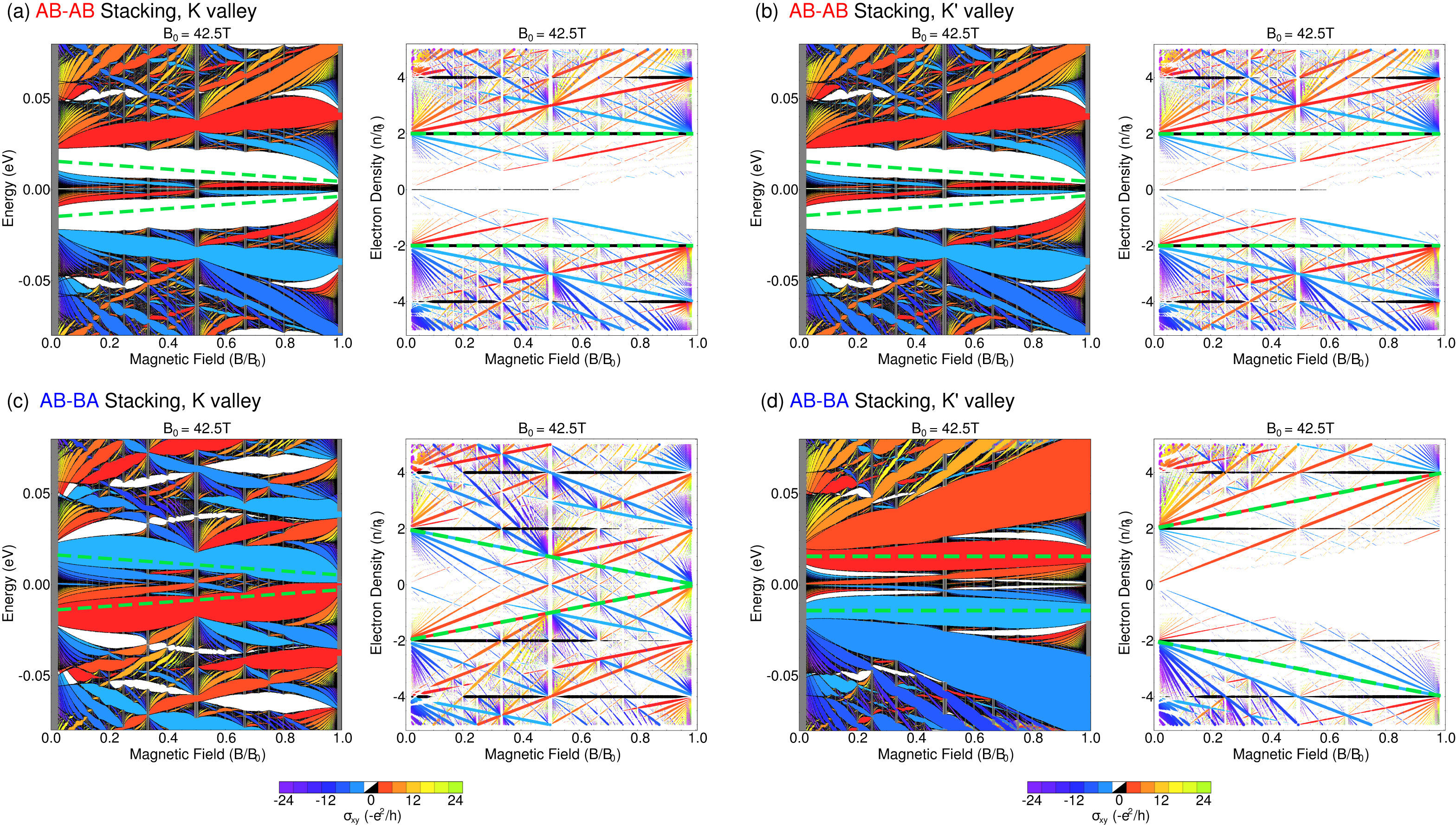}
	\caption{(Colour online) (a) and (b): Energy spectrum (left panel) and Wannier diagram (right panel) of the $K$ (a) and $K'$ (b) valley of the AB-AB stacked TDBG in a minimal model. (c) and (d): Plots similar to (a) and (b), but for the AB-BA stacked TDBG. The dashed lines show the correspondence of the gaps in between the energy spectrum and Wannier diagram.}
	\label{fig_s1}
\end{figure*}

As noted in Sec.\,\ref{sec:energy_spectrum} and Fig.\,\ref{fig2},
in the minimal model,
the Chern numbers
at the gaps above and below
the central bands near the charge neutrality point are zero in the AB-AB stacked TDBG
but is finite (two) in the AB-BA stacked TDBG.
In the AB-AB stacking, as a result,
the electron density between the gaps in both the $K$ and $K'$ valleys
remain the same throughout the entire range of magnetic fields [Figs.\,\ref{fig_s1}(a) and (b)].
In the AB-BA stacked TDBG, on the contrary,
the electron density between the gaps decreases in $K$ while it increases in $K'$
as the magnetic field strength increases [Figs.\,\ref{fig_s1}(c) and (d)].

\bibliography{TDBG}

\end{document}